\documentclass[letter]{aa} % for the letters 
\usepackage{amsmath}
\usepackage{multirow}
\usepackage{graphicx}
\usepackage{gensymb}
\usepackage{natbib}
\usepackage[colorlinks,citecolor=blue,linkcolor=blue,urlcolor=blue,bookmarks=false,hypertexnames=true]{hyperref}
\usepackage{lscape} 
\usepackage{adjustbox}
\usepackage[normalem]{ulem}    
\usepackage{tablefootnote}

\newcommand{\nustar}{\textit{NuSTAR }}

\newcommand{\swift}{{\it Swift }}

\usepackage{graphicx}
\usepackage{txfonts}
\begin{document} 

\title{Universality of coronal properties in accreting black holes across mass and accretion rate}
%Universality of BH coronal properties
\author{
Sudip Chakraborty\inst{1},
Ajay Ratheesh\inst{2},
Francesco Tombesi\inst{3,4,5,6,7},
Rodrigo Nemmen\inst{8,9}, and
Srimanta Banerjee\inst{10}}

\institute{
Universit\'e Paris Saclay, Universit\'e Paris Cit\'e, CEA, CNRS, AIM, F-91191 Gif-sur-Yvette, France.\\
              \email{sudip.chakraborty@cea.fr}
\and
INAF-IAPS, Via del Fosso del Cavaliere 100, I-00133 Rome, Italy. 
\and
Department of Physics, University of Rome 'Tor Vergata', via della Ricerca Scientifica 1, 00133, Rome, Italy.
\and
INAF - Astronomical Observatory of Rome, Via Frascati 33, I-00078 Monte Porzio Catone (Rome), Italy
\and
Department of Astronomy, University of Maryland, College Park, MD 20742, USA
\and
NASA/Goddard Space Flight Center, Greenbelt, MD 20771, USA
\and
INFN - Rome Tor Vergata, via della Ricerca Scientifica 1, 00133, Rome, Italy
\and
Universidade de S\~ao Paulo, Instituto de Astronomia, Geof\'{\i}sica e Ci\^encias Atmosf\'ericas, Departamento de Astronomia, S\~ao Paulo, SP 05508-090, Brazil
\and
Kavli Institute for Particle Astrophysics and Cosmology (KIPAC), Stanford University, Stanford, CA 94305, USA
\and
Inter University Centre for Astronomy and Astrophysics, Post bag 4, Ganeshkhind, Pune, India
}

\date{}

% \abstract{}{}{}{}{} 
% 5 {} token are mandatory
 
  \abstract
  % context heading (optional)
  % {} leave it empty if necessary  
   {}
  % aims heading (mandatory)
   {Through their radio loudness, lack of thermal UV emission from the accretion disk and power-law dominated spectra, Low Luminosity AGN (LLAGN) display similarity with the hard state of stellar-mass black hole X-Ray Binaries (BHBs). In this work we perform a systematic hard X-ray spectral study of a carefully selected sample of unobscured LLAGN using archival \nustar data, to understand the central engine properties in the lower accretion regime. }
  % methods heading (mandatory)
   {We analyze the \nustar spectra of a sample of 16 LLAGN. We model the continuum emission with detailed Comptonization models. }
  % results heading (mandatory)
   {We find a strong anti-correlation between the optical depth and the electron temperature of the corona,  previously also observed in the brighter Seyferts. This anti-correlation is present irrespective of the shape of the corona, and the slope of this anti-correlation in the log space for LLAGN (0.68-1.06) closely matches that of the higher accretion rate Seyferts (0.55-1.11) and hard state of BHBs ($\sim$0.87). This anti-correlation may indicate a departure from a fixed disk-corona configuration in radiative balance. }
  % conclusions heading (optional), leave it empty if necessary 
   {Our result, therefore, demonstrates a possible universality in Comptonization processes of black hole X-ray sources across multiple orders of magnitude in mass and accretion rate.
}

\titlerunning{Universality of BH coronal properties}
\authorrunning{S. Chakraborty et al.}
   \maketitle
%
%-------------------------------------------------------------------

\section{Introduction} \label{sec:introduction}
Low-Luminosity Active Galactic Nuclei (LLAGN) are intrinsically faint sub-Eddington accreting systems, with average bolometric luminosities $L_{\rm  bol}\sim10^{38-43}\,{\rm erg\,s^{-1}}$ and average Eddington ratio $\lambda_{\rm Edd}=L_{\rm bol}/L_{\rm Edd}\sim10^{-2}-10^{-5}$ \citep{Ho_2009}, $L_{\rm Edd}$ being the Eddington luminosity, as opposed to their luminous AGN counterparts \citep[$\lambda_{\rm Edd}\sim0.1-1$;][]{Kollmeier_2006}. It is believed that the bulk of the supermassive black hole (SMBH) population in the local universe reside in the form of underfed LLAGN \citep{Ho_1995, Ho_2008}, making up the major fraction of the lifetime of SMBHs \citep{Martini_2004,Shin_2010}, but contributing little to their growth.

Observationally, LLAGN differ from the luminous AGN: they lack significant X-ray variability \citep{Pellegrini_2000,Ho_2008}, their broadband SEDs (Spectral Energy Distributions) have no quasar-like ``big blue bump'' in the UV continuum \citep{Quataert_1999} \citep[an indicator of the standard optically thick, geometrically thin accretion disk,][]{Ho_1999, Ho_2008, Nemmen_2006}, they typically have weak or non-existent, narrow ${\rm Fe\,K\alpha}$ emission line \citep{Terashima_2002}  along with the occasional presence of broad double-peaked ${\rm H\alpha}$ lines \citep{Storchi-Bergmann_2003} - all consistent with the absence of a thin accretion disk or the presence of a truncated thin accretion disk \citep[truncated at $\gtrsim100GM/c^{2}$,][]{Chen_1989}. Furthermore, all of these observational evidences hint towards the LLAGN accretion mode likely being Advection-Dominated Accretion Flows \citep[ADAF; ][]{Narayan_RIAF_1998,Nemmen_2014}, in which the hot, geometrically thick, optically thin accretion flow has typically low radiative efficiencies ($L\ll0.1\dot{M}c^{2}$) and low accretion rate ($\dot{M}\lesssim0.01\dot{M}_{\rm Edd}$). Relativistic jets are also thought to play an important role throughout the LLAGN broadband SED \citep{Pian_2010, Nemmen_2014,Fernandez_2023}. Furthermore, LLAGNs do not follow some of the correlations established in their higher luminosity counterparts \citep[e.g., instead of the positive $\Gamma$ vs $\lambda_{\rm Edd}$ correlation in brighter AGNs \citep{Sobolewska_2009}, LLAGNs show an anti-correlation between the two parameters;][]{Gu_2009,Yang_2015,She_2018}.

The X-ray spectrum of AGN is usually dominated by a primary power-law component, attributed to a hot plasma \citep[monikered the `corona'; ][]{Haardt_1991,Haardt_1997} Compton up-scattering the optical/UV photons emitted by the underlying accretion disk into the X-ray band. A high-energy cut-off in the hard X-ray spectrum, indicative of the temperature of the corona, is one of the important signatures of this Comptonization process \citep{Balokovic_2015,Brenneman_2014}. Additionally, reflection features, in the form of an iron line complex around $\sim$6.4 keV as well as a ``Compton hump" at $\sim$20--40 keV, are sometimes present. The simplest description of Comptonizing coronae is obtained by measuring the photon power-law index and the cut-off in the hard X-ray spectrum. The former depends on the interplay between the electron temperature and the optical depth, whereas the latter is directly related to the electron temperature of the corona. While in brighter AGN, the primary source of X-ray photons is considered as the unsaturated Comptonization of thermal photons from a  standard geometrically thin, optically thick accretion disk \citep{Shakura_1973}, for LLAGN the contribution might also originate from the synchrotron self-Compton emission. Therefore, to compare and understand the coronal properties between the brighter AGN and the LLAGN, hard X-ray observations are of paramount importance.

While the brightest AGN have been studied extensively in the past with hard X-ray  satellites, such as BeppoSAX \citep{Dadina_2007}, INTEGRAL \citep{Molina_2013} and \swift-BAT \citep{Ricci_2017}, hard X-ray class studies of LLAGN is sparse \citep[e.g.,][]{Diaz_2023}. The Nuclear Spectroscopic Telescope Array \citep[\nustar,][]{Harrison_NuSTAR_2013}, the first focusing X-ray telescope at hard X-rays, allows us to systematically study the hard X-ray signatures of LLAGN in unprecedented details thanks to its focusing optics, broad and high-quality spectral coverage between 3 to 79 keV. Therefore, \nustar is suitable for studying the hard X-ray spectra of AGN with high
sensitivity, discriminating between the primary X-ray emission and the reflected component. Alone, or with simultaneous observations with other X-ray observatories operating below 10 keV, such as XMM-Newton, Suzaku and \swift-XRT, it has provided strong constraints on the coronal properties of many bright AGN \citep{Brenneman_2014,Fabian_2015,Matt_2015,Fabian_2017,Tortosa_2018}.

Recently, \citet{Tortosa_2018} studied the coronal properties of a sample of bright Seyferts using \nustar spectra. Along with the previously reported correlations, they have found clear indications of unexplained anti-correlation between the electron temperature and the optical depth. To explore the validity of these correlations at much lower luminosities, a systematic Comptonization study of LLAGN with \nustar has to be carried out. In our present work, we aim to bridge this gap. In section~\ref{sec:selection}, we discuss our sample selection. In section~\ref{sec:analysis}, we explore the selected sample of LLAGN with different spectral models and find the correlations between the important parameters. We present the key results in section~\ref{sec:results}. Finally, in section~\ref{sec:discussion}, we discuss the implications of this study.

\section{Data selection} \label{sec:selection}
The sample considered in this work was selected predominantly from Palomar sample \citep{Saikia_2018, Nagar_2005}, and supplemented with sources from the BASS DR1\footnote{\url{https://www.bass-survey.com/dr1.html}} \citep{Ricci_2017}, as well as the following works: \citet{Nemmen_2014, Kawamuro_2016, Ho_2009, Hernandez-Garcia_2016, Gonzalez-Martin_2012, Eracleous_2010, Terashima_2002, Ursini_2015}. Since the main motive of this work is to study the properties of the central engine in these sources, we identified unobscured, Compton-thin LLAGN by imposing upper constraint of $10^{24} \rm \ cm^{-2}$ on the hydrogen equivalent column density ($N_{\rm H}$) and $\sim 10^{43}$ erg/s on the bolometric luminosity. All the sources and the corresponding \nustar data used in this work are outlined in Table. \ref{tab:llagn_details}. Methods of reduction of \nustar data is given in Appendix \ref{appendix:data_reduction}. We also impose a minimum count-rate threshold of 4$\times$10$^{-2}$ cts/s in the 3-79 keV energy range in both FPMA and FPMB to reach a satisfactory signal-to-noise level for spectral analysis. This produces the 16 sources, described in table~\ref{tab:llagn_details}, constituting our sample. For details on the uniform reduction of the $NuSTAR$ data, see Appendix~\ref{appendix:data_reduction}.

\section{Spectral Analysis} \label{sec:analysis}
The spectral fitting and statistical analysis are carried out using the XSPEC version v-12.12.0 \citep{Arnaud_1996}. To jointly fit FPMA and FPMB, a cross-normalization constant (\textsc{Constant} model in XSPEC) is allowed to vary freely for FPMB and is assumed to be unity for FPMA. We also restrict the energy ranges of the individual datasets to 3-25, 3-50 or 3-79 keV, based on the quality of the data. All the models, as described below, include the Galactic absorption through the implementation of the \textsc{tbabs} model. The corresponding abundances are set as per the solar abundances in \citet{Wilms_2000}. The neutral hydrogen column densities ($N_{\rm H}$) are fixed to values found in the literature (see table~\ref{tab:llagn_details})  for all the described models. All parameter uncertainties are reported at the $1\sigma$ confidence level for one parameter of interest.

To get a detailed understanding of the Comptonization processes and to directly compare with previous studies of other accreting black hole systems, we require reliable estimates of the optical depth ($\tau$) and electron temperatures ($kT_{\rm e}$) of the coronae. To directly fit the optical depths and electron temperatures in the 3-79 keV \nustar data, we use the Comptonization  model \textsc{compTT} \citep{Titarchuk_1994} in XSPEC. \textsc{compTT} models the thermal Comptonization emission of a hot plasma cooled by soft photons with a Wien law distribution and includes special relativistic effects. This model is valid both in the case of optically thick and optically thin plasma. The Comptonized spectrum is determined completely by the plasma temperature and the so-called $\beta$ parameter which is independent of geometry. We use both the available slab and sphere geometry in this work. For the former, we set the geometry switch to 0.5, and for the latter, we set it to 2. In both cases, the corresponding $\beta$ values are calculated from optical depth using analytic approximations \citep{Titarchuk_1994}. We use the redshift values mentioned in table~\ref{tab:llagn_details} and assume the seed black-body temperature to be fixed at 10 eV \citep{Younes_2019}. In the cases where excess at Fe-K energies (around 6.4 keV) are observed, we include a Gaussian emission line (\textsc{gauss} in XSPEC). Whenever the lines are found to be too narrow to be resolved by \nustar, we freeze the width of the Gaussian line ($\sigma$) to zero. The details of the best-fit continuum parameters are presented in table~\ref{tab:params}, while emission lines are described in table~\ref{tab:gauss}. The best-fit parameters and $\chi^2/\rm{d.o.f.}$ values stated in table~\ref{tab:params} are corrected for reflection features with physically motivated models as well, as described in the next paragraph. From table~\ref{tab:params}, it can be seen that both the geometries of \textsc{compTT} model result in statistically similar fits. Physical constraints over $kT_{\rm e}$ are found for 12 and 11 cases for the slab and the sphere geometries, respectively. Out of these, the values of $kT_{\rm e}$ for NGC 4258 are found to be unphysical ($<10$ keV). 

Of the 10 LLAGN with prominent iron lines, 4 are found to have central energies of $\ge 6.4$ keV, and 6 are found to have line energies $\le 6.4$ keV. Upon further investigations on the available literature of these 10 LLAGN, the iron lines in the latter 6 were found to have AGN origin, and the former 6 are found to have originated from hot diffused gas. Most early-type galaxies emit an extended hot diffuse X-ray component usually fitted with an emission model from an optically thin plasma \citep{Fabbiano_1989}. To account for this, we use an APEC component (Astrophysical Plasma Emission Code, \citet{Smith_2001}) along with the \textsc{compTT} continuum to model instead of the Gaussian. We assume solar metallicity and fit for the plasma temperature. For the LLAGN with AGN-origin Fe-K lines, and to account for the occasional reflection hump noted in a few of these sources, we use the \textsc{pexmon} model \citep{Nandra_2007} for reflection from a neutral medium. The model component is dependent on the inclination angle of the source, which we fix to 45$^{\circ}$, and on elemental abundances which we assume to be solar. Compared to other similar models such as \textsc{pexrav}, \textsc{pexmon} has the advantage of self-consistently including reflection due to atomic species such as the Fe K$\alpha$, Fe K$\beta$ and Ni K$\alpha$ \citep{Nandra_2007}. The inclusion of \textsc{apec} and \textsc{pexmon} improves the fits in all the cases. Finally, in the case of NGC 5506, broad iron lines accompanied by a Compton hump at 30-50 keV, are observed. To test the possibility of relativistic broadening, we convolve the \textsc{pexmon} model with smeared relativistic  accretion  disk  line  profiles  using \textsc{relconv} \citep{Dauser_2010}. Hereafter, we use the \textsc{apec}/\textsc{pexmon}-corrected $kT_{\rm e}$ values for studying correlations between the parameters to make them more robust. 

\section{Results} \label{sec:results}
\begin{figure}[htb!]
\centering
\includegraphics[width=\linewidth]{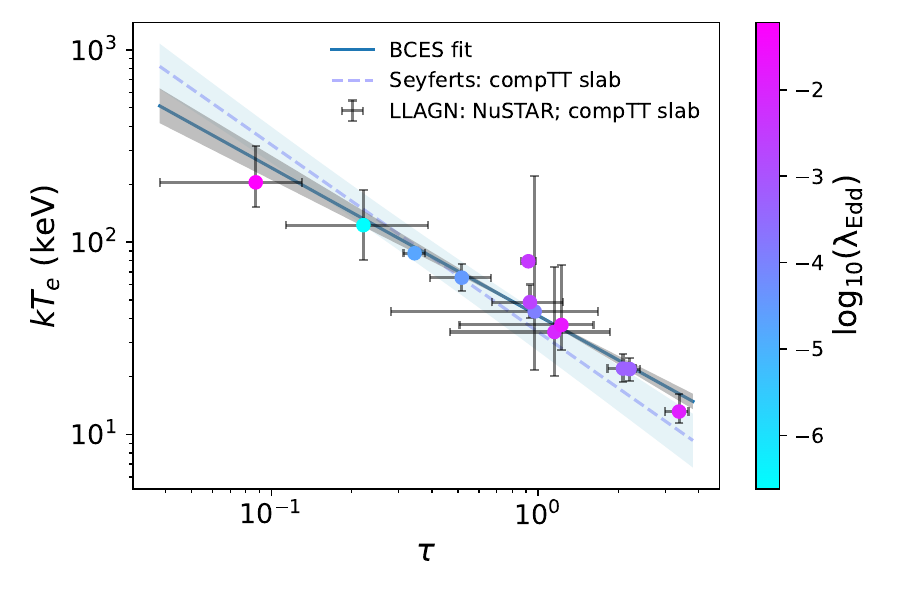}
\includegraphics[width=\linewidth]{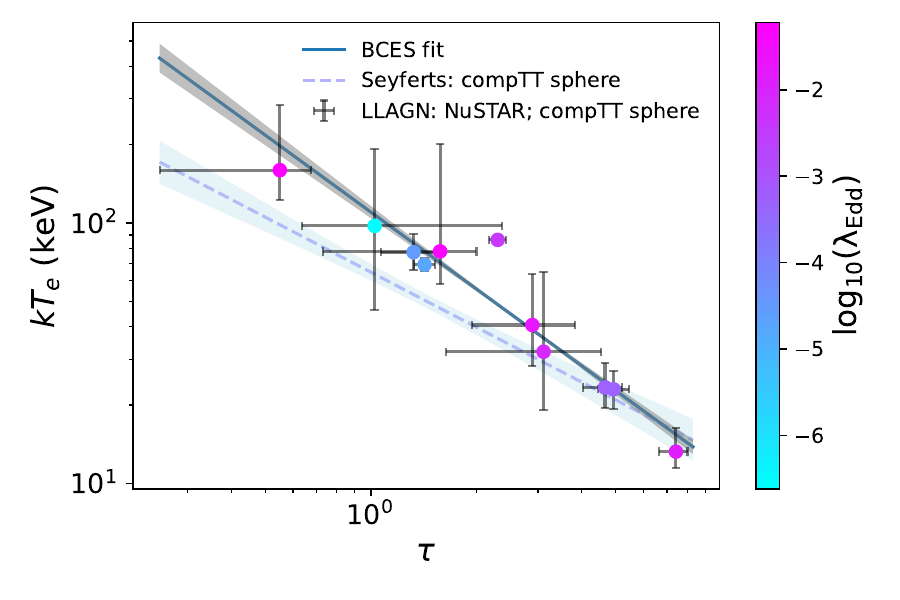}
\caption{The electron temperature ($kT_{\rm e}$) vs optical depth ($\tau$) for different geometries of the corona from the CompTT model. Top panel: slab geometry, Bottom panel: spherical geometry. In both the panels, the best-fit parameters are shown along with their 1$\sigma$ errorbars. The data points have been color-coded as per their Eddington ratio $\lambda_{\rm Edd}$.
The solid lines in both panels indicate the BCES best-fit lines to our LLAGN sample and the shaded grey area contains their confidence regions. For easier comparison, the best-fit line of the Seyferts \citep{Tortosa_2018} has been placed for the corresponding geometry in both panels; the BCES best-fit line is indicated with dashed blue and the corresponding confidence region in light-blue. For details about the fit, refer to section~\ref{sec:results}.}
\label{fig:sphereslab}
\end{figure}

From figure~\ref{fig:sphereslab}, we observe an anti-correlation between $\tau$ and $kT_{\rm e}$ in our sample of LLAGN, for both the slab and spherical geometry of the corona.
This anti-correlation is similar to the one found by \citet{Tortosa_2018} for the more luminous AGN. 

We find the Spearman's rank correlation coefficients between $\tau$ and $k T_{\rm e}$ (see Appendix~\ref{appendix:spearman} for details) for the LLAGN to be $\rho=-0.75 \pm 0.14$ for the spherical corona and $\rho=-0.81 \pm 0.11$ for the slab corona, with corresponding $p$-values of 0.003 and 0.002. The histogram of the $\rho$ values for both the geometries are displayed in figure~\ref{fig:spearman}. Such high negative values of $\rho$ suggest strong anti-correlation between $kT_{\rm e}$ and $\tau$ for both the coronal geometries, although the anti-correlation is found to be stronger for the slab geometry.

In order to directly perform a regression analysis in the $kT_{\rm e}$ vs $\tau$ plane for our LLAGN sample, we implement the bi-variate correlated errors and intrinsic scatter (BCES\footnote{\url{https://github.com/rsnemmen/BCES}}), following the prescription of \citet{Akritas_1996}. We perform $5\times 10^4$ trials and adopt the orthogonal least squares BCES line as our best fit.
Fitting the $kT_{\rm e}$ and $\tau$ in the logarithmic space ($\log (kT_{\rm e})=a\log (\tau)+b$), we find the best-fit parameters to be:
\begin{equation*}
    \text{For slab geometry:}\\
    a=-0.77 \pm 0.09; \; b=1.62 \pm 0.01
\end{equation*}
\begin{equation*}
    \text{For spherical geometry:}\\
    a=-0.98 \pm 0.08; \; b=2.04 \pm 0.03
\end{equation*}
To compare these values directly with the ones for luminous AGN, we perform the same BCES-based fit on the $kT_{\rm e}$ vs $\tau$ data from \citet{Tortosa_2018}. For such bright AGN, we find the best-fit values to be:
\begin{equation*}
    \text{For slab geometry:}\\
    a=-0.97 \pm 0.14; \; b=1.53 \pm 0.15
\end{equation*}
\begin{equation*}
    \text{For spherical geometry:}\\
    a=-0.70 \pm 0.15; \; b=1.81 \pm 0.05
\end{equation*}
The $kT_{\rm e}$ vs $\tau$ plots along with the BCES best-fit lines and the corresponding confidence regions for the two geometries, are displayed in figure~\ref{fig:sphereslab}.

\begin{figure}[htb!]
\centering
\includegraphics[width=\linewidth]{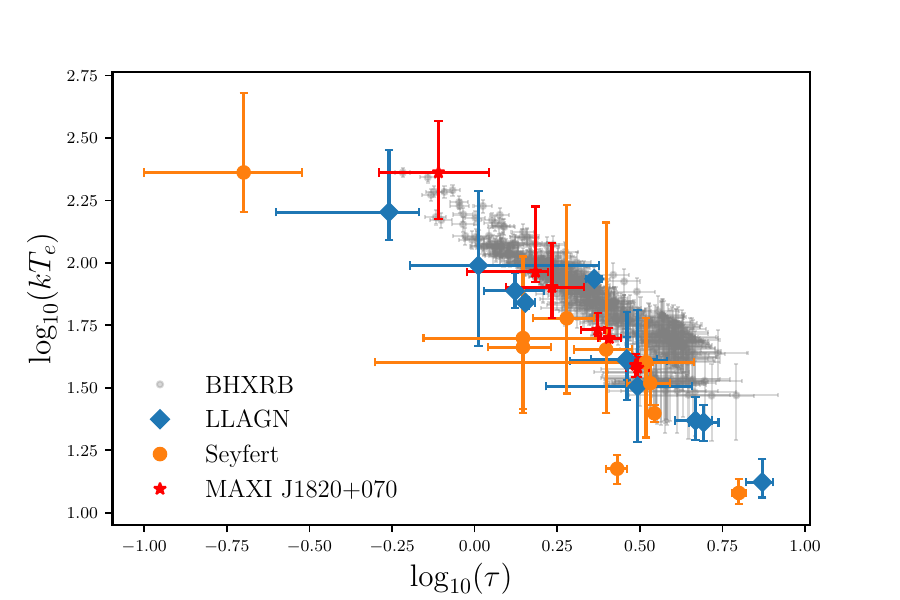}
\caption{The electron temperature ($kT_{\rm e}$) vs optical depth ($\tau$) for the spherical geometry of the corona across different class of objects: LLAGN (green), Seyferts from \citep{Tortosa_2018} (orange). For a direct comparison with the Galactic BHBs, the hard state data (with \textsc{compPS} model fit) from \citet{Banerjee_2020} has been presented in light grey points. Furthermore, to indicate the evolution of BHBs along the presented anti-correlation, hard state data points for MAXI J1820+070 \citep{Buisson_2019} have been plotted in red. For a discussion on the implication of this universality, refer to section~\ref{sec:discussion}.}
\label{fig:allsources}
\end{figure}

\section{Discussions and conclusions} \label{sec:discussion}
In this work, we probe the coronal properties of a sample of 16 carefully selected unobscured LLAGN with \nustar. 
{We analyze the \nustar spectra using slab and sphere geometries of \textsc{compTT}, a model of thermal Comptonization of the seed disk photons with an assumed seed photon temperature of 10 eV. Assuming a simple parallel with Black Hole X-ray binary hard state (see the later discussions in this section), this would imply an inner edge of accretion disks at $\sim 100-500 \ R_{\rm g}$. This is roughly in line with what we find for NGC 5506 (Appendix \ref{appendix:NGC 5506}). Although the seed photons are assumed to originate from the disk, they could also be of any other origin without considerably affecting the hard X-ray spectra, as any potential changes would primarily be observed on the lower energy side of the spectrum.} We also investigate the iron line complex for each LLAGN, and account for the emission lines using either a diffused plasma emission using \textsc{apec}, or reflection from neutral material via \textsc{pexmon}. Using a uniform analysis over the entire sample, we derive robust values of $kT_{\rm e}$ and $\tau$ of the coronae. We find a similar anti-correlation between the optical depth and the electron temperatures for LLAGN, to that found in the more luminous AGN \citep{Tortosa_2018}. This anti-correlation, depicted in figure~\ref{fig:sphereslab}, is found to hold true for both the slab and spherical geometry of the corona. Overall, the slab geometry is found to have more significant anti-correlation compared to the spherical geometry. The best-fit slope for LLAGN ($-0.76\pm0.09$ for slab and $-0.98\pm0.08$ for sphere) closely follows the slope for brighter Seyferts ($-0.98\pm0.18$ for slab and $0.70\pm0.16$ for sphere).

It has been suggested for LLAGN in the ADAF regime that the scattering optical depth is related to the thickness and density of the accretion flow. This would imply that, to first order, $\tau$ is related to the mass accretion rate as $\tau = 0.03 (\dot{m}/10^{-3})$ assuming a radiatively inefficient accretion flow \citep{Narayan_1995}, where $\dot{m}$ is the accretion rate in units of the Eddington rate ($\dot{M}_{\rm Edd} \equiv L_{\rm Edd}/(0.1 c^2)$). Taking the best-fit values found from the observations for the slab geometry, this would suggest that for LLAGNs, the electron temperature depends relatively strongly on $\dot{m}$ as $k T_e = 599 (\dot{m}/10^{-3})^{-0.77}$ keV or $k T_e = 10^{9.8} (\dot{m}/10^{-3})^{-0.77}$ K. However, no such correlation is evident between $k T_e$ and $\dot{m}$ or $\tau$ and $\dot{m}$ from figure \ref{fig:sphereslab}, although our methodology assumes a standard Comptonizing corona instead of an ADAF in the first place. For example, the Spearman's rank coefficient between $\tau$ and $\lambda_{\rm Edd}$ is only $0.20 \pm 0.16$ and $0.24 \pm 0.14$ for spherical and slab geometry, respectively; while between $k T_e$ and $\lambda_{\rm Edd}$ it is only $-0.11 \pm 0.16$ and $-0.18 \pm 0.15$ for spherical and slab geometry, respectively. This suggests that in this picture, the Comptonizing region in LLAGNs may be completely uncorrelated with the ADAF region and could instead resemble the Comptonizing coronae found in more luminous AGNs.

Remarkably, the same anti-correlation ($\rho=-0.84 \pm 0.01$) between $kT_{\rm e}$ and $\tau$ has been found for the hard state spectra of BHBs \citep{Banerjee_2020}. A linear regression analysis in logarithmic space yields a strikingly similar slope ($-0.87\pm0.02$). As the BHB data-set in \citet{Banerjee_2020} uses spherical geometry of \textsc{compPS} model, this can be compared with the spherical geometries of our LLAGN sample or the bright AGN sample of \citet{Tortosa_2018}. For the BHB sample, however, the intercept is found to be slightly higher ($2.17 \pm 0.01$). This systematic difference between BHBs and AGN would indicate that for any given $\tau$ value, the spectra of BHBs are harder \citep{Middei_2019} than their supermassive counterparts. This presence of similar anti-correlation and the remarkable similarity in Comptonization parameters across different classes of accreting black hole sources hints towards a universality of the coronal physics in all of them.

The anti-correlation between $kT_{\rm e}$ and $\tau$ indicates a departure from a fixed disk-corona configuration in radiative balance \citep{Tortosa_2018}. The invalidation of a fixed disk-corona configuration can possibly occur due to a change in the coronal geometry. For example, a reduction of coronal height (in the lamppost configuration) would imply a larger Compton cooling from the disk (with an increase of $\tau$) and, thereby, a smaller $kT_{\rm e}$. Such a variation in the coronal geometry was earlier proposed to explain the evolution of spectral and timing features of the BHB MAXI J1820+070 in the hard state \citep{kara2018,Buisson_2019}. And, as can be seen from the red points in figure~\ref{fig:allsources}, MAXI J1820+070 also occupies a similar parameter space along the $kT_{\rm e}-\tau$ anti-correlation line as it evolves from its hard state and its corona contracts. On the other hand, the violation of radiative balance due to a change in the disk fraction (i.e., the fraction of intrinsic disk emission to the total flux) for a fixed disk-corona system can also expound this anti-correlation \citep{Tortosa_2018}. The disk fraction can alter if the inner edge of the accretion disk evolves as the supply of the soft seed disk photons to the corona depend upon this. Thus, as the disk recedes, the disk fraction falls, and the average number of scattering of thermal seed photons with coronal electrons decreases (i.e., optical depth drops). This leads to a decrease in Compton cooling and, thereby, an increase in $kT_{\rm e}$. Hence for a highly truncated disk, $\tau$ is expected to be lower and $kT_{\rm e}$ to be higher. In our work, we find, in all three iterations of our blurred reflection model \textsc{relconv$\otimes$pexmon} implementation that the $R_{\rm in}$ for NGC 5506 is significantly large, indicating a highly truncated disk (see Appendix~\ref{appendix:NGC 5506}). While the disk truncation radii for other LLAGN are not confirmed, there are indications that LLAGN, as a population, might occupy the hard state branch  \citep[usually associated with large disk truncation, e.g.][]{Done_2007} in the `q'-diagram \citep{Fernandez_2023}.

\clearpage

\bibliographystyle{aa}
\bibliography{llagn_nustar}

\begin{thebibliography}{72}
\expandafter\ifx\csname natexlab\endcsname\relax\def\natexlab#1{#1}\fi

\bibitem[{{Akritas} \& {Bershady}(1996)}]{Akritas_1996}
{Akritas}, M.~G. \& {Bershady}, M.~A. 1996, \apj, 470, 706

\bibitem[{{Ananna} {et~al.}(2022{\natexlab{a}}){Ananna}, {Urry}, {Ricci},
  {Natarajan}, {Hickox}, {Trakhtenbrot}, {Treister}, {Weigel}, {Ueda}, {Koss},
  {Bauer}, {Temple}, {Balokovi{\'c}}, {Mushotzky}, {Auge}, {Sanders}, {Kakkad},
  {Sartori}, {Marchesi}, {Harrison}, {Stern}, {Oh}, {Caglar}, {Powell},
  {Podjed}, \& {Mej{\'\i}a-Restrepo}}]{Tasnim_2022a}
{Ananna}, T.~T., {Urry}, C.~M., {Ricci}, C., {et~al.} 2022{\natexlab{a}},
  \apjl, 939, L13

\bibitem[{{Ananna} {et~al.}(2022{\natexlab{b}}){Ananna}, {Weigel},
  {Trakhtenbrot}, {Koss}, {Urry}, {Ricci}, {Hickox}, {Treister}, {Bauer},
  {Ueda}, {Mushotzky}, {Ricci}, {Oh}, {Mej{\'\i}a-Restrepo}, {den Brok},
  {Stern}, {Powell}, {Caglar}, {Ichikawa}, {Wong}, {Harrison}, \&
  {Schawinski}}]{Tasnim_2022b}
{Ananna}, T.~T., {Weigel}, A.~K., {Trakhtenbrot}, B., {et~al.}
  2022{\natexlab{b}}, \apjs, 261, 9

\bibitem[{{Arnaud}(1996)}]{Arnaud_1996}
{Arnaud}, K.~A. 1996, Astronomical Society of the Pacific Conference Series,
  Vol. 101, {XSPEC: The First Ten Years}, ed. G.~H. {Jacoby} \& J.~{Barnes}, 17

\bibitem[{{Balokovi{\'c}} {et~al.}(2015){Balokovi{\'c}}, {Matt}, {Harrison},
  {Zoghbi}, {Ballantyne}, {Boggs}, {Christensen}, {Craig}, {Esmerian},
  {Fabian}, {F{\"u}rst}, {Hailey}, {Marinucci}, {Parker}, {Reynolds}, {Stern},
  {Walton}, \& {Zhang}}]{Balokovic_2015}
{Balokovi{\'c}}, M., {Matt}, G., {Harrison}, F.~A., {et~al.} 2015, \apj, 800,
  62

\bibitem[{{Banerjee} {et~al.}(2020){Banerjee}, {Gilfanov}, {Bhattacharyya}, \&
  {Sunyaev}}]{Banerjee_2020}
{Banerjee}, S., {Gilfanov}, M., {Bhattacharyya}, S., \& {Sunyaev}, R. 2020,
  \mnras, 498, 5353

\bibitem[{{Brenneman} {et~al.}(2014){Brenneman}, {Madejski}, {Fuerst}, {Matt},
  {Elvis}, {Harrison}, {Ballantyne}, {Boggs}, {Christensen}, {Craig}, {Fabian},
  {Grefenstette}, {Hailey}, {Madsen}, {Marinucci}, {Rivers}, {Stern}, {Walton},
  \& {Zhang}}]{Brenneman_2014}
{Brenneman}, L.~W., {Madejski}, G., {Fuerst}, F., {et~al.} 2014, \apj, 788, 61

\bibitem[{{Buisson} {et~al.}(2019){Buisson}, {Fabian}, {Barret}, {F{\"u}rst},
  {Gandhi}, {Garc{\'\i}a}, {Kara}, {Madsen}, {Miller}, {Parker}, {Shaw},
  {Tomsick}, \& {Walton}}]{Buisson_2019}
{Buisson}, D.~J.~K., {Fabian}, A.~C., {Barret}, D., {et~al.} 2019, \mnras, 490,
  1350

\bibitem[{{Chen} {et~al.}(1989){Chen}, {Halpern}, \& {Filippenko}}]{Chen_1989}
{Chen}, K., {Halpern}, J.~P., \& {Filippenko}, A.~V. 1989, \apj, 339, 742

\bibitem[{{Dadina}(2007)}]{Dadina_2007}
{Dadina}, M. 2007, \aap, 461, 1209

\bibitem[{{Dauser} {et~al.}(2010){Dauser}, {Wilms}, {Reynolds}, \&
  {Brenneman}}]{Dauser_2010}
{Dauser}, T., {Wilms}, J., {Reynolds}, C.~S., \& {Brenneman}, L.~W. 2010,
  \mnras, 409, 1534

\bibitem[{{Diaz} {et~al.}(2023){Diaz}, {Hern{\`a}ndez-Garc{\'\i}a},
  {Ar{\'e}valo}, {L{\'o}pez-Navas}, {Ricci}, {Koss}, {Gonzalez-Martin},
  {Balokovi{\'c}}, {Osorio-Clavijo}, {Garc{\'\i}a}, \& {Malizia}}]{Diaz_2023}
{Diaz}, Y., {Hern{\`a}ndez-Garc{\'\i}a}, L., {Ar{\'e}valo}, P., {et~al.} 2023,
  \aap, 669, A114

\bibitem[{{Done} {et~al.}(2007){Done}, {Gierli{\'n}ski}, \&
  {Kubota}}]{Done_2007}
{Done}, C., {Gierli{\'n}ski}, M., \& {Kubota}, A. 2007, \aapr, 15, 1

\bibitem[{{Eracleous} {et~al.}(2010){Eracleous}, {Hwang}, \&
  {Flohic}}]{Eracleous_2010}
{Eracleous}, M., {Hwang}, J.~A., \& {Flohic}, H. M.~L.~G. 2010, \apj, 711, 796

\bibitem[{{Fabbiano}(1989)}]{Fabbiano_1989}
{Fabbiano}, G. 1989, \araa, 27, 87

\bibitem[{{Fabian} {et~al.}(2017){Fabian}, {Lohfink}, {Belmont}, {Malzac}, \&
  {Coppi}}]{Fabian_2017}
{Fabian}, A.~C., {Lohfink}, A., {Belmont}, R., {Malzac}, J., \& {Coppi}, P.
  2017, \mnras, 467, 2566

\bibitem[{{Fabian} {et~al.}(2015){Fabian}, {Lohfink}, {Kara}, {Parker},
  {Vasudevan}, \& {Reynolds}}]{Fabian_2015}
{Fabian}, A.~C., {Lohfink}, A., {Kara}, E., {et~al.} 2015, \mnras, 451, 4375

\bibitem[{{Fern{\'a}ndez-Ontiveros} {et~al.}(2023){Fern{\'a}ndez-Ontiveros},
  {L{\'o}pez-L{\'o}pez}, \& {Prieto}}]{Fernandez_2023}
{Fern{\'a}ndez-Ontiveros}, J.~A., {L{\'o}pez-L{\'o}pez}, X., \& {Prieto}, A.
  2023, \aap, 670, A22

\bibitem[{{Gonz{\'a}lez-Mart{\'\i}n} \& {Vaughan}(2012)}]{Gonzalez-Martin_2012}
{Gonz{\'a}lez-Mart{\'\i}n}, O. \& {Vaughan}, S. 2012, \aap, 544, A80

\bibitem[{{Gu} \& {Cao}(2009)}]{Gu_2009}
{Gu}, M. \& {Cao}, X. 2009, \mnras, 399, 349

\bibitem[{{Haardt} \& {Maraschi}(1991)}]{Haardt_1991}
{Haardt}, F. \& {Maraschi}, L. 1991, \apjl, 380, L51

\bibitem[{{Haardt} {et~al.}(1997){Haardt}, {Maraschi}, \&
  {Ghisellini}}]{Haardt_1997}
{Haardt}, F., {Maraschi}, L., \& {Ghisellini}, G. 1997, \apj, 476, 620

\bibitem[{{Harrison} {et~al.}(2013){Harrison}, {Craig}, {Christensen},
  {Hailey}, {Zhang}, {Boggs}, {Stern}, {Cook}, {Forster}, {Giommi},
  {Grefenstette}, {Kim}, {Kitaguchi}, {Koglin}, {Madsen}, {Mao}, {Miyasaka},
  {Mori}, {Perri}, {Pivovaroff}, {Puccetti}, {Rana}, {Westergaard}, {Willis},
  {Zoglauer}, {An}, {Bachetti}, {Barri{\`e}re}, {Bellm}, {Bhalerao},
  {Brejnholt}, {Fuerst}, {Liebe}, {Markwardt}, {Nynka}, {Vogel}, {Walton},
  {Wik}, {Alexander}, {Cominsky}, {Hornschemeier}, {Hornstrup}, {Kaspi},
  {Madejski}, {Matt}, {Molendi}, {Smith}, {Tomsick}, {Ajello}, {Ballantyne},
  {Balokovi{\'c}}, {Barret}, {Bauer}, {Blandford}, {Brandt}, {Brenneman},
  {Chiang}, {Chakrabarty}, {Chenevez}, {Comastri}, {Dufour}, {Elvis}, {Fabian},
  {Farrah}, {Fryer}, {Gotthelf}, {Grindlay}, {Helfand}, {Krivonos}, {Meier},
  {Miller}, {Natalucci}, {Ogle}, {Ofek}, {Ptak}, {Reynolds}, {Rigby},
  {Tagliaferri}, {Thorsett}, {Treister}, \& {Urry}}]{Harrison_NuSTAR_2013}
{Harrison}, F.~A., {Craig}, W.~W., {Christensen}, F.~E., {et~al.} 2013, \apj,
  770, 103

\bibitem[{{Hern{\'a}ndez-Garc{\'\i}a}
  {et~al.}(2016){Hern{\'a}ndez-Garc{\'\i}a}, {Masegosa},
  {Gonz{\'a}lez-Mart{\'\i}n}, {M{\'a}rquez}, \&
  {Perea}}]{Hernandez-Garcia_2016}
{Hern{\'a}ndez-Garc{\'\i}a}, L., {Masegosa}, J., {Gonz{\'a}lez-Mart{\'\i}n},
  O., {M{\'a}rquez}, I., \& {Perea}, J. 2016, \apj, 824, 7

\bibitem[{{Ho}(1999)}]{Ho_1999}
{Ho}, L.~C. 1999, \apj, 516, 672

\bibitem[{{Ho}(2008)}]{Ho_2008}
{Ho}, L.~C. 2008, \araa, 46, 475

\bibitem[{{Ho}(2009)}]{Ho_2009}
{Ho}, L.~C. 2009, \apj, 699, 626

\bibitem[{{Ho} {et~al.}(1995){Ho}, {Filippenko}, \& {Sargent}}]{Ho_1995}
{Ho}, L.~C., {Filippenko}, A.~V., \& {Sargent}, W.~L. 1995, \apjs, 98, 477

\bibitem[{{Jana} {et~al.}(2023){Jana}, {Chatterjee}, {Chang}, {Nandi},
  {Rubinur}, {Kumari}, {Naik}, {Safi-Harb}, \& {Ricci}}]{Jana_2023}
{Jana}, A., {Chatterjee}, A., {Chang}, H.-K., {et~al.} 2023, \mnras
  [\eprint[arXiv]{2307.07966}]

\bibitem[{{Joye} \& {Mandel}(2003)}]{DS9_2003ASPC..295..489J}
{Joye}, W.~A. \& {Mandel}, E. 2003, in Astronomical Society of the Pacific
  Conference Series, Vol. 295, Astronomical Data Analysis Software and Systems
  XII, ed. H.~E. {Payne}, R.~I. {Jedrzejewski}, \& R.~N. {Hook}, 489

\bibitem[{{Kara} {et~al.}(2019){Kara}, {Steiner}, {Fabian}, {Cackett},
  {Uttley}, {Remillard}, {Gendreau}, {Arzoumanian}, {Altamirano}, {Eikenberry},
  {Enoto}, {Homan}, {Neilsen}, \& {Stevens}}]{kara2018}
{Kara}, E., {Steiner}, J.~F., {Fabian}, A.~C., {et~al.} 2019, \nat, 565, 198

\bibitem[{{Kawamuro} {et~al.}(2016){Kawamuro}, {Ueda}, {Tazaki}, {Terashima},
  \& {Mushotzky}}]{Kawamuro_2016}
{Kawamuro}, T., {Ueda}, Y., {Tazaki}, F., {Terashima}, Y., \& {Mushotzky}, R.
  2016, arXiv e-prints, arXiv:1604.07915

\bibitem[{{Kollmeier} {et~al.}(2006){Kollmeier}, {Onken}, {Kochanek}, {Gould},
  {Weinberg}, {Dietrich}, {Cool}, {Dey}, {Eisenstein}, {Jannuzi}, {Le Floc'h},
  \& {Stern}}]{Kollmeier_2006}
{Kollmeier}, J.~A., {Onken}, C.~A., {Kochanek}, C.~S., {et~al.} 2006, \apj,
  648, 128

\bibitem[{{Marinucci} {et~al.}(2018){Marinucci}, {Bianchi}, {Braito}, {Matt},
  {Nardini}, \& {Reeves}}]{Marinucci_2018}
{Marinucci}, A., {Bianchi}, S., {Braito}, V., {et~al.} 2018, \mnras, 478, 5638

\bibitem[{{Martini}(2004)}]{Martini_2004}
{Martini}, P. 2004, in Coevolution of Black Holes and Galaxies, ed. L.~C. {Ho},
  169

\bibitem[{{Masini} {et~al.}(2022){Masini}, {Wijesekera}, {Celotti}, \&
  {Boorman}}]{Masini_2022}
{Masini}, A., {Wijesekera}, J.~V., {Celotti}, A., \& {Boorman}, P.~G. 2022,
  \aap, 663, A87

\bibitem[{{Matt} {et~al.}(2015){Matt}, {Balokovi{\'c}}, {Marinucci},
  {Ballantyne}, {Boggs}, {Christensen}, {Comastri}, {Craig}, {Gandhi},
  {Hailey}, {Harrison}, {Madejski}, {Madsen}, {Stern}, \& {Zhang}}]{Matt_2015}
{Matt}, G., {Balokovi{\'c}}, M., {Marinucci}, A., {et~al.} 2015, \mnras, 447,
  3029

\bibitem[{{Mehdipour} {et~al.}(2021){Mehdipour}, {Kriss}, {Kaastra}, {Wang},
  {Mao}, {Costantini}, {Arav}, {Behar}, {Bianchi}, {Branduardi-Raymont},
  {Brotherton}, {Cappi}, {De Marco}, {Di Gesu}, {Ebrero}, {Grafton-Waters},
  {Kaspi}, {Matt}, {Paltani}, {Petrucci}, {Pinto}, {Ponti}, {Ursini}, \&
  {Walton}}]{Mehdipour_2021}
{Mehdipour}, M., {Kriss}, G.~A., {Kaastra}, J.~S., {et~al.} 2021, \aap, 652,
  A150

\bibitem[{{Middei} {et~al.}(2019){Middei}, {Bianchi}, {Marinucci}, {Matt},
  {Petrucci}, {Tamborra}, \& {Tortosa}}]{Middei_2019}
{Middei}, R., {Bianchi}, S., {Marinucci}, A., {et~al.} 2019, \aap, 630, A131

\bibitem[{{Middei} {et~al.}(2022){Middei}, {Marinucci}, {Braito}, {Bianchi},
  {De Marco}, {Luminari}, {Matt}, {Nardini}, {Perri}, {Reeves}, \&
  {Vagnetti}}]{Middei_2022}
{Middei}, R., {Marinucci}, A., {Braito}, V., {et~al.} 2022, \mnras, 514, 2974

\bibitem[{{Molina} {et~al.}(2013){Molina}, {Bassani}, {Malizia}, {Stephen},
  {Bird}, {Bazzano}, \& {Ubertini}}]{Molina_2013}
{Molina}, M., {Bassani}, L., {Malizia}, A., {et~al.} 2013, \mnras, 433, 1687

\bibitem[{{Nagar} {et~al.}(2005){Nagar}, {Falcke}, \& {Wilson}}]{Nagar_2005}
{Nagar}, N.~M., {Falcke}, H., \& {Wilson}, A.~S. 2005, \aap, 435, 521

\bibitem[{{Nandra} {et~al.}(2007){Nandra}, {O'Neill}, {George}, \&
  {Reeves}}]{Nandra_2007}
{Nandra}, K., {O'Neill}, P.~M., {George}, I.~M., \& {Reeves}, J.~N. 2007,
  \mnras, 382, 194

\bibitem[{{Narayan} {et~al.}(1998){Narayan}, {Mahadevan}, \&
  {Quataert}}]{Narayan_RIAF_1998}
{Narayan}, R., {Mahadevan}, R., \& {Quataert}, E. 1998, in Theory of Black Hole
  Accretion Disks, ed. M.~A. {Abramowicz}, G.~{Bj{\"o}rnsson}, \& J.~E.
  {Pringle}, 148--182

\bibitem[{{Narayan} \& {Yi}(1995)}]{Narayan_1995}
{Narayan}, R. \& {Yi}, I. 1995, \apj, 452, 710

\bibitem[{{Nasa High Energy Astrophysics Science Archive Research Center
  (Heasarc)}(2014)}]{ftools_2014ascl.soft08004N}
{Nasa High Energy Astrophysics Science Archive Research Center (Heasarc)}.
  2014, {HEAsoft: Unified Release of FTOOLS and XANADU}

\bibitem[{{Nemmen} {et~al.}(2014){Nemmen}, {Storchi-Bergmann}, \&
  {Eracleous}}]{Nemmen_2014}
{Nemmen}, R.~S., {Storchi-Bergmann}, T., \& {Eracleous}, M. 2014, \mnras, 438,
  2804

\bibitem[{{Nemmen} {et~al.}(2006){Nemmen}, {Storchi-Bergmann}, {Yuan},
  {Eracleous}, {Terashima}, \& {Wilson}}]{Nemmen_2006}
{Nemmen}, R.~S., {Storchi-Bergmann}, T., {Yuan}, F., {et~al.} 2006, \apj, 643,
  652

\bibitem[{{Osorio-Clavijo} {et~al.}(2020){Osorio-Clavijo},
  {Gonz{\'a}lez-Mart{\'\i}n}, {Papadakis}, {Masegosa}, \&
  {Hern{\'a}ndez-Garc{\'\i}a}}]{Osorio_2020}
{Osorio-Clavijo}, N., {Gonz{\'a}lez-Mart{\'\i}n}, O., {Papadakis}, I.~E.,
  {Masegosa}, J., \& {Hern{\'a}ndez-Garc{\'\i}a}, L. 2020, \mnras, 491, 29

\bibitem[{{Pal} {et~al.}(2022){Pal}, {Stalin}, {Mallick}, \& {Rani}}]{Pal_2022}
{Pal}, I., {Stalin}, C.~S., {Mallick}, L., \& {Rani}, P. 2022, \aap, 662, A78

\bibitem[{{Pellegrini} {et~al.}(2000){Pellegrini}, {Cappi}, {Bassani}, {della
  Ceca}, \& {Palumbo}}]{Pellegrini_2000}
{Pellegrini}, S., {Cappi}, M., {Bassani}, L., {della Ceca}, R., \& {Palumbo},
  G.~G.~C. 2000, \aap, 360, 878

\bibitem[{{Pian} {et~al.}(2010){Pian}, {Romano}, {Maoz}, {Cucchiara}, {Pagani},
  \& {Parola}}]{Pian_2010}
{Pian}, E., {Romano}, P., {Maoz}, D., {et~al.} 2010, \mnras, 401, 677

\bibitem[{Prieto {et~al.}(2016)Prieto, Fernández-Ontiveros, Markoff, Espada,
  \& González-Martín}]{Prieto_2016}
Prieto, M.~A., Fernández-Ontiveros, J.~A., Markoff, S., Espada, D., \&
  González-Martín, O. 2016, Monthly Notices of the Royal Astronomical
  Society, 457, 3801

\bibitem[{{Quataert} {et~al.}(1999){Quataert}, {Di Matteo}, {Narayan}, \&
  {Ho}}]{Quataert_1999}
{Quataert}, E., {Di Matteo}, T., {Narayan}, R., \& {Ho}, L.~C. 1999, \apjl,
  525, L89

\bibitem[{{Rani} {et~al.}(2017){Rani}, {Stalin}, \& {Rakshit}}]{Rani_2017}
{Rani}, P., {Stalin}, C.~S., \& {Rakshit}, S. 2017, \mnras, 466, 3309

\bibitem[{{Ricci} {et~al.}(2017){Ricci}, {Trakhtenbrot}, {Koss}, {Ueda}, {Del
  Vecchio}, {Treister}, {Schawinski}, {Paltani}, {Oh}, {Lamperti}, {Berney},
  {Gandhi}, {Ichikawa}, {Bauer}, {Ho}, {Asmus}, {Beckmann}, {Soldi},
  {Balokovi{\'c}}, {Gehrels}, \& {Markwardt}}]{Ricci_2017}
{Ricci}, C., {Trakhtenbrot}, B., {Koss}, M.~J., {et~al.} 2017, \apjs, 233, 17

\bibitem[{{Saikia} {et~al.}(2018){Saikia}, {K{\"o}rding}, {Coppejans},
  {Falcke}, {Williams}, {Baldi}, {Mchardy}, \& {Beswick}}]{Saikia_2018}
{Saikia}, P., {K{\"o}rding}, E., {Coppejans}, D.~L., {et~al.} 2018, \aap, 616,
  A152

\bibitem[{{Shakura} \& {Sunyaev}(1973)}]{Shakura_1973}
{Shakura}, N.~I. \& {Sunyaev}, R.~A. 1973, \aap, 500, 33

\bibitem[{{She} {et~al.}(2018){She}, {Ho}, {Feng}, \& {Cui}}]{She_2018}
{She}, R., {Ho}, L.~C., {Feng}, H., \& {Cui}, C. 2018, \apj, 859, 152

\bibitem[{{Shin} {et~al.}(2010){Shin}, {Ostriker}, \& {Ciotti}}]{Shin_2010}
{Shin}, M.-S., {Ostriker}, J.~P., \& {Ciotti}, L. 2010, \apj, 711, 268

\bibitem[{{Smith} {et~al.}(2001){Smith}, {Brickhouse}, {Liedahl}, \&
  {Raymond}}]{Smith_2001}
{Smith}, R.~K., {Brickhouse}, N.~S., {Liedahl}, D.~A., \& {Raymond}, J.~C.
  2001, \apjl, 556, L91

\bibitem[{{Sobolewska} \& {Papadakis}(2009)}]{Sobolewska_2009}
{Sobolewska}, M.~A. \& {Papadakis}, I.~E. 2009, \mnras, 399, 1597

\bibitem[{{Storchi-Bergmann} {et~al.}(2003){Storchi-Bergmann}, {Nemmen da
  Silva}, {Eracleous}, {Halpern}, {Wilson}, {Filippenko}, {Ruiz}, {Smith}, \&
  {Nagar}}]{Storchi-Bergmann_2003}
{Storchi-Bergmann}, T., {Nemmen da Silva}, R., {Eracleous}, M., {et~al.} 2003,
  \apj, 598, 956

\bibitem[{{Terashima} {et~al.}(2002){Terashima}, {Iyomoto}, {Ho}, \&
  {Ptak}}]{Terashima_2002}
{Terashima}, Y., {Iyomoto}, N., {Ho}, L.~C., \& {Ptak}, A.~F. 2002, \apjs, 139,
  1

\bibitem[{{Titarchuk}(1994)}]{Titarchuk_1994}
{Titarchuk}, L. 1994, \apj, 434, 570

\bibitem[{{Tortosa} {et~al.}(2018){Tortosa}, {Bianchi}, {Marinucci}, {Matt}, \&
  {Petrucci}}]{Tortosa_2018}
{Tortosa}, A., {Bianchi}, S., {Marinucci}, A., {Matt}, G., \& {Petrucci}, P.~O.
  2018, \aap, 614, A37

\bibitem[{{Ursini} {et~al.}(2015){Ursini}, {Marinucci}, {Matt}, {Bianchi},
  {Tortosa}, {Stern}, {Ar{\'e}valo}, {Ballantyne}, {Bauer}, {Fabian},
  {Harrison}, {Lohfink}, {Reynolds}, \& {Walton}}]{Ursini_2015}
{Ursini}, F., {Marinucci}, A., {Matt}, G., {et~al.} 2015, \mnras, 452, 3266

\bibitem[{{Wilms} {et~al.}(2000){Wilms}, {Allen}, \& {McCray}}]{Wilms_2000}
{Wilms}, J., {Allen}, A., \& {McCray}, R. 2000, \apj, 542, 914

\bibitem[{Wong {et~al.}(2017)Wong, Nemmen, Irwin, \& Lin}]{Wong_2017}
Wong, K.-W., Nemmen, R.~S., Irwin, J.~A., \& Lin, D. 2017, The Astrophysical
  Journal, 849, L17

\bibitem[{{Yang} {et~al.}(2015){Yang}, {Xie}, {Yuan}, {Zdziarski},
  {Gierli{\'n}ski}, {Ho}, \& {Yu}}]{Yang_2015}
{Yang}, Q.-X., {Xie}, F.-G., {Yuan}, F., {et~al.} 2015, \mnras, 447, 1692

\bibitem[{{Younes} {et~al.}(2019){Younes}, {Ptak}, {Ho}, {Xie}, {Terasima},
  {Yuan}, {Huppenkothen}, \& {Yukita}}]{Younes_2019}
{Younes}, G., {Ptak}, A., {Ho}, L.~C., {et~al.} 2019, \apj, 870, 73

\bibitem[{Young {et~al.}(2018)Young, McHardy, Emmanoulopoulos, \&
  Connolly}]{Young_2018}
Young, A.~J., McHardy, I., Emmanoulopoulos, D., \& Connolly, S. 2018, Monthly
  Notices of the Royal Astronomical Society, 476, 5698–5703

\end{thebibliography}

\begin{acknowledgements}
We thank the anonymous reviewer for the constructive comments in improving this manuscript. This work made use of data from the NuSTAR mission, a project led by the California Institute of Technology, managed by the Jet Propulsion Laboratory, and funded by the National Aeronautics and Space Administration. We thank the NuSTAR Operations, Software and Calibration teams for their support with the execution and analysis of these observations. This research has also made use of the \nustar Data Analysis Software (NuSTARDAS), jointly developed by the ASI Science Data Center (ASDC, Italy) and the California Institute of Technology (USA). This project has received funding from the European Union's Horizon 2020 research and innovation programme under grant agreement n°101004168, the XMM2ATHENA project. The authors thank Prof. A. R. Rao and Prof. Sudip Bhattacharyya for their constructive suggestions and comments for improving this work, and Dr. Tonima Tasnim Ananna for the BASS DR2 estimates. RN acknowledges support by FAPESP (Funda\c{c}\~ao de Amparo \`a Pesquisa do Estado de S\~ao Paulo) under grants 2017/01461-2 and 2022/10460-8. We thank the anonymous referee for the constructive comments.
\end{acknowledgements}

\appendix

\section{Data reduction}\label{appendix:data_reduction}
The \nustar data used in this work are reduced using v2.1.1 of the NuSTARDAS pipeline using the \nustar CALDB v20211020. The \textit{nupipeline} tool was used to generate cleaned level2 event files. We used DS9 \citep{DS9_2003ASPC..295..489J}, to select the source and background region. The source region was selected manually to include the maximum source contribution and the background was selected away from the source to avoid contamination, both using $1^{\prime}$ circular regions. The \textit{nuproducts} tool was then used to extract the background subtracted source spectrum. For non-variable sources with multiple data sets, we did a combined fit by combining spectra from those observations using the \textit{addspec} (version 1.4.0) tool of the FTOOLS \citep{ftools_2014ascl.soft08004N}. Before combining the spectra, we checked the source variabilities by probing the consistency between the fluxes and the photon indices for \textsc{cutoffpl} model fits of the spectra of different epochs. All the spectra were then binned using \textit{grppha} tool of the FTOOLS to obtain at least 25 counts per bin, in order to use $\chi^2$-statistics.

\section{Distribution of Spearman's rank correlation coefficient}\label{appendix:spearman}

\begin{figure}[htb!]
\centering
\includegraphics[width=\linewidth]{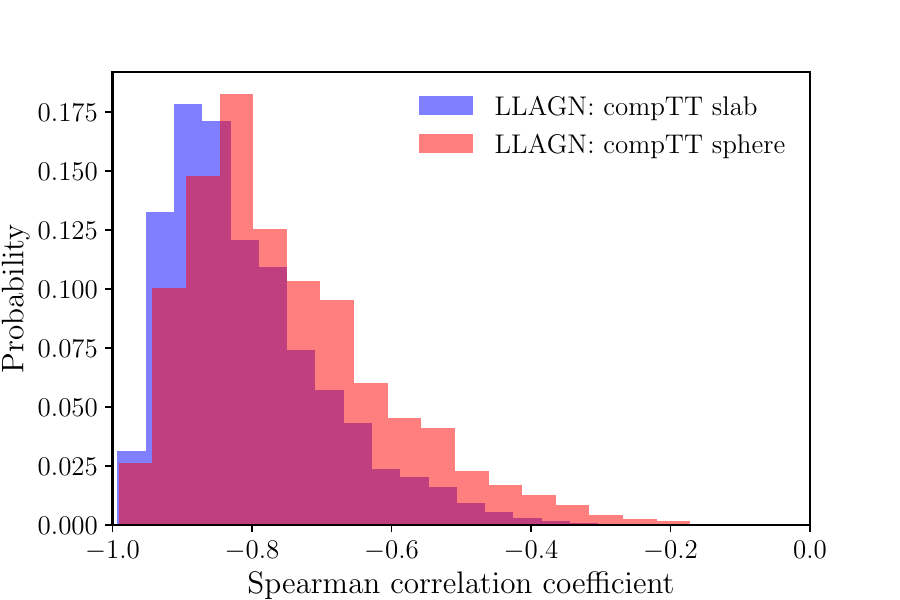}
\caption{The histograms of the Spearman rank correlation coefficients ($\rho$) from the simulated data for the slab (blue) and spherical (red) geometries of the corona in LLAGN. While the histogram  of slab geometry shows a more prominent peak at a higher $\rho$ value than the spherical geometry, high negative mean $\rho$ for both the geometries indicate moderate to significant anti-correlation between $kT_{\rm e}$ and $\tau$ present in both. For details about the histograms, refer to section~\ref{sec:results}.}
\label{fig:spearman}
\end{figure}

To check the goodness of our anti-correlation in the presence of error bars in both $\tau$ and $k T_{\rm e}$, we perform a Monte Carlo calculation by randomly generating simulated data points for each of the sources by using the best-fit parameter and the covariance matrix, repeating the procedure $2 \times 10^4$ times. We then calculate Spearman's rank correlation coefficient for each of the trials and note the mean and standard deviation of the resulting distribution. We calculate the $p$-values by calculating the number of times $\rho_{\rm sim} \ge \rho_{\rm obs}$ (where $\rho_{\rm sim}$ and $\rho_{\rm obs}$ are the Spearman's rank coefficient derived from the simulated in consideration and the corresponding value for the actual coefficient calculated from the data points, respectively), and dividing it by the total number of simulations. The distribution of $\rho_{\rm sim}$ for the slab and spherical geometry is portrayed in figure~\ref{fig:spearman} and the relevant values are stated and discussed in section~\ref{sec:results}.

\section{Further details about specific sources: NGC 5506}\label{appendix:NGC 5506}
For NGC 5506, \textsc{compTT+(relconv$\otimes$pexmon)} resulted in a much better and more consistent fit than \textsc{compTT+pexmon}. While the simple \textsc{pexmon} implementation resulted in a $\chi^2/\rm{d.o.f.}$ of 2057/1827, assuming the inclination to be fixed at $45^{\circ}$ and freezing the spin parameter $a$ to its maximum allowed value of 0.998, the \textsc{relconv$\otimes$pexmon} implementation gives a much better fit with a $\chi^2/\rm{d.o.f.}$ of 1989/1826. The best-fit $kT_{\rm e}$ is found to be $>36.1$ keV and the inner edge of accretion disk $R_{\rm in}>303 \ R_{\rm g}$ (where $R_{\rm g}$ is the gravitational radius of the central black hole). Changing the inclination to $30^{\circ}$ results in an even better fit, with a $\chi^2/\rm{d.o.f.}$ of 1956/1826, the best-fit $kT_{\rm e}=78^{+131}_{-27}$ keV and $R_{\rm in}=126^{+26}_{-23} \ R_{\rm g}$. Finally, letting the inclination vary freely, results in the best fit of the three cases, with a $\chi^2/\rm{d.o.f.}$ of 1937/1825. This best-fit result is the one presented in table~\ref{tab:params} and used for the correlation study. The corresponding  inclination and  $R_{\rm in}$ are found to be $18.3^{+1.5}_{-1.3}$ degrees and $58^{+10}_{-8} \ R_{\rm g}$, respectively.

\section{Additional tables}\label{appendix:tables}

\begin{table*}[h]
\begin{center}
\resizebox{\textwidth}{!}{%
\begin{tabular}{ c c c c c c c c c c} 
 \hline
 Name & Obs ID & Date & Exposure  & \nustar & Spectral & log(Mass)  & $z$ & $N_{\rm H}$ & {log ($\lambda_{Edd}$)} \\ 
 & & (yyyy-mm-dd) & (ks) & ref. & type & log($M_{\odot}$) & & $\rm (\times 10^{22}cm^{-2})$ & \\
 \hline
  M81 (NGC 3031) & 60101049002 & 2015-05-18 & 209 & \citet{Young_2018} & S1.5/L1.5 & 7.9 &	0.001 &	1.86 & $-6.62^a$  \\
 [1ex]
 M87 (NGC 4486) & 60201016002 & 2017-02-15  & 50.0  & \citet{Wong_2017} & L2 & 9.5 & 0.004 &	0.40 & $-5.44^d$\\
                & 60466002002 & 2018-04-24  & 21.1  & This work &  &  & & & \\
                & 90202052002 & 2017-04-11  & 24.4  & \citet{Wong_2017} &  & & & & \\
                & 90202052004 & 2017-04-14  & 22.5  & \citet{Wong_2017} &  & & & & \\
 [1ex]
 NGC 1052 & 60061027002 & 2013-02-14 & 15.5 & \citet{Rani_2017} & L1.9 & 8.7 &	0.005 &	12.78  & $-3.94^b$ \\
  & 60201056002 & 2017-01-17 & 59.7 & \citet{Osorio_2020} &  &  & & &  \\
  & 90701624002 & 2021-07-17 & 20.5 & This work &  &  & & &\\
  [1ex]
  NGC 2655 & 60160341002 & 2016-11-02 & 0.51  & \citet{Diaz_2023} & S2 & 8.2 &	0.005 & 4.27 & $-3.87^b$ \\
& 60160341004 & 2016-11-10 & 16.0 & This work & & & & & \\
[1ex]
NGC 2992$^*$ & 60160371002 & 2015-12-02 & 20.8 & \citet{Marinucci_2018} & S1.9/1.5 & 8.3 & 0.00771 & 0.78 & $-3.30^b$ \\
 & 90501623002 & 2019-05-10 & 57.4 & \citet{Middei_2022} &  &  &  & & \\
[1ex]
 NGC 3227 & 60202002004 & 2016-11-25 & 42.0& \citet{Pal_2022} & L1.9 & 6.7 & 0.004 & 0.09 & $-1.70^c$\\
& 60202002006 & 2016-11-29 & 39.7 & \citet{Pal_2022} & & &\\
& 60202002008 & 2016-12-01 & 41.7 & \citet{Pal_2022} & & &\\
& 60202002010 & 2016-12-05 & 40.9 & \citet{Pal_2022} & & &\\
& 60202002012 & 2016-12-09 & 39.3 & \citet{Pal_2022} & & &\\
& 60202002014 & 2017-01-21 & 47.6 & \citet{Pal_2022} & & &\\
[1ex]
 NGC 3718 & 60301031002 & 2017-10-24 & 24.5 & \citet{Diaz_2023} & L1.9 & 8.1 &	0.003 &	0.14 & $-4.49^b$ \\
          & 60301031004 & 2017-10-27 & 90.4 & \citet{Diaz_2023} & & & & &\\
          & 60301031006 & 2017-10-30 & 57.4 & \citet{Diaz_2023} & & & & & \\
          & 60301031008 & 2017-11-03 & 57.0 & \citet{Diaz_2023} & & & & &\\
[1ex]
NGC 3998 & 60201050002 & 2016-10-25 & 104 & \citet{Younes_2019} & L1.9 & 8.9 & 0.003 & 0.03 & $-4.47^b$ \\ 
[1ex]
 NGC 4258 & 60101046002 & 2015-11-16 & 54.8 & \citet{Masini_2022}  & S1.9 & 7.6 &	0.002	& 8.0 & $-4.28^b$	 \\
 & 60101046004 & 2016-01-10 & 103.6 & \citet{Masini_2022} &  &  & & 	&  \\
 [1ex]
 NGC 4395 & 60061322002 & 2013-05-10 &19.2 & \citet{Rani_2017} & S1.8 & 5.4 &	0.001 &	0.04 & $-1.90^a$ \\
[1ex]
 NGC 4579 & 60201051002 & 2016-12-06 & 117 & \citet{Younes_2019} & S1.9/L1.9 & 8.1	& 0.004 &	0.03 & $-3.564^b$ \\
          & 60201056002 & 2017-01-17 & 59.7 & \citet{Osorio_2020} & & &\\
 [1ex]
 NGC 5033 & 60601023002 & 2020-12-08 & 103.5 & \citet{Diaz_2023} & S1.9 &7.7  & 0.002 & 0.01 & $-4.00^b$\\
 & 60601023004 & 2020-12-12 & 53.3 & \citet{Diaz_2023} &  &  &  & & \\
[1ex]
 NGC 5273 & 60061350002 & 2014-07-14 & 21.1 & \citet{Rani_2017} & S1.5 & 6.7 & 0.003 & 0.07  & $-2.43^a$\\
  & 90801618002 & 2022-07-03 & 18.0 & This work & & & & &\\
[1ex]
 NGC 5290 & 60160554002 & 2021-07-28 & 18.9  &  \citet{Jana_2023} & S2 & 7.8 &	0.009	& 0.91	& $-2.64^a$ \\
[1ex]
 NGC 5506 & 60061323002 & 2014-04-01 & 57.0  & \citet{Matt_2015} & S1.9 & 6.0 &	0.006 &	0.31 & $-1.22^a$ \\
  & 60501015002 & 2019-12-28 & 61.4  & This work &  &  &	& &	\\
   & 60501015004 & 2020-02-09 & 47.5  & This work &  &  &	 &	& \\
 [1ex]

 NGC 7213 & 60001031002 & 2014-10-05 & 109 & \citet{Ursini_2015} & S1.5 & 7.1 &	0.005 &	0.02 & $-2.08^a$ \\
 [1ex]
%  Pictor A & 60101047002 & 2015-12-03 & 109  &  \citet{Zhao_2021} & L1.5 & 7.6 &	0.034	& 0.002	& $-1.34^a$ \\
 % [1ex]

\hline
\end{tabular}
}
\caption{The key properties of all the sources in our LLAGN sample and the details of its corresponding \nustar observation used in this work. For the source with $^*$, the different observations were analyzed separately. The spectral types are abbreviated as S: Seyfert, L: LINER. $z$, $N_{\rm H}$ and $\lambda_{\rm Edd}$ denote redshift, absorbing column density and Eddington ratio, respectively.  {$^a$:\cite{Tasnim_2022a, Tasnim_2022b}, $^b$: \cite{Jana_2023}, $^c$:\cite{Mehdipour_2021}, $^d$:\cite{Prieto_2016}}}
\label{tab:llagn_details}
\end{center}
\end{table*}

\begin{table*}[htb!]
\begin{center}
\caption{Best-fit parameters of \textsc{compTT} models for both slab and spherical geometries in our sample of LLAGN. The corresponding errorbars are stated at 1$\sigma$ levels. Here $kT_{\rm e}$ is the electron temperature of the corona and $\tau$ is its optical depth. For all the applicable LLAGN, the best-fit model includes, in addition to \textsc{compTT}, suitable \textsc{apec} and \textsc{pexmon} models. See section~\ref{sec:analysis} for a detailed discussion.}
\begin{tabular}{ c c c c c c c c  }
\hline
 & \multicolumn{3}{c}{Sphere} & \multicolumn{3}{c}{Slab} &   \\ 
\hline \\ LLAGN name	& $kT_{\rm e}$ & $\tau$ & $\chi^2$/d.o.f & kT$_e$ & $\tau$ & $\chi^2$/d.o.f & 3-30 keV Flux \\ 
 & (keV) &  &  & (keV) &  & & ($\times 10^{-11}$ ergs cm$^{-2}$ s$^{-1}$)\\ \hline
M81 & $97.5^{+95.7}_{-51.1}$ & $1.03^{+1.35}_{-0.39}$ &  1277/1232 & $122.6^{+64.2}_{-41.6}$ & $0.22^{+0.17}_{-0.11}$ &  1277/1232  & 3.40 \\ 
[1ex]
M87 & $144.3^{+3.6}_{-40.5}$ & $<0.02$ &  433.64/292 & $83.2^{+1.5}_{-17.4}$ & $<0.02$ &  435.78/292 & 0.87\\ 
[1ex]
NGC1052 & $>218.9$ & $0.27^{+0.43}_{-0.01}$ &  980/842 & $19.2^{+6.1}_{-2.7}$ & $2.33^{+0.27}_{-0.22}$ &  982/842 & 1.55 \\ 
[1ex]
NGC 2655 & $62.5^{+43.9}_{-24.2}$ & $0.79^{+0.68}_{-0.46}$ &  65.41/50 & $49.1^{+50.3}_{-16.5}$ & $0.32^{+0.27}_{-0.24}$ &  65.58/50 & 0.65 \\ 
[1ex]
NGC 2992 & $23.4^{+5.7}_{-3.9}$ & $4.66^{+0.56}_{-0.62}$ &  871.60/940 & $22.0^{+4.3}_{-3.2}$ & $2.08^{+0.24}_{-0.26}$ &  872.85/940 & 12.07 \\ 
[1ex]
 & $23.0^{+4.1}_{-3.7}$ & $4.93^{+0.54}_{-0.47}$ &  1457.46/1467 & $21.9^{+3.0}_{-2.8}$ & $2.2^{+0.21}_{-0.19}$ &  1458.21/1467 & 17.31 \\
[1ex]
NGC 3227 & $40.6^{+23.1}_{-12.3}$ & $2.89^{+0.93}_{-0.95}$ &  1901/1674 & $37.1^{+38.6}_{-9.7}$ & $1.22^{+0.39}_{-0.72}$ &  1900/1674 & 7.56 \\ 
[1ex]
NGC 3718 & $77.2^{+13.8}_{-11.4}$ & $1.33^{+0.3}_{-0.26}$ &  209.03/199 & $65.3^{+11.8}_{-9.6}$ & $0.52^{+0.15}_{-0.12}$ &  209.14/199 & 0.22 \\ 
[1ex]
NGC 3998 & $69.3^{+3.9}_{-3.7}$ & $1.42^{+0.1}_{-0.1}$ &  584.01/613 & $87.6^{+5.0}_{-4.7}$ & $0.34^{+0.03}_{-0.03}$ &  584.01/613 & 1.20 \\ 
[1ex]
NGC 4258 & $8.1^{+1.4}_{-0.9}$ & $7.93^{+0.58}_{-0.72}$ &  607/567 & $8.0^{+1.5}_{-0.8}$ & $3.63^{+0.27}_{-0.36}$ &  607/567 &  0.54 \\ 
[1ex]
NGC 4395 & $137.4^{+198.9}_{-56.8}$ & $2.38^{+1.14}_{-1.23}$ &  356.63/317 & $130.8^{+152.9}_{-52.3}$ & $0.92^{+0.54}_{-0.5}$ &  355.96/317 & 2.07 \\ 
[1ex]
NGC 4579 & $>84.4$ & $0.12^{+1.08}_{-0.0}$ &  652/668 & $>105.9$ & $0.02^{+0.26}_{-0.0}$ &  652/668 & 1.20 \\ 
[1ex]
NGC 5033 & $36.1^{+198.7}_{-19.1}$ & $3.33^{+2.16}_{-2.76}$ &  935.25/842 & $26.2^{+254.7}_{-9.3}$ & $1.75^{+0.66}_{-1.39}$ &  935.28/842 & 1.44\\ 
[1ex]
NGC 5273 & $86.2^{+3.5}_{-3.3}$ & $2.3^{+0.13}_{-0.12}$ &  752.19/721 & $79.5^{+3.1}_{-3.0}$ & $0.92^{+0.06}_{-0.06}$ &  752.78/721 & 3.77 \\ 
[1ex]
NGC 5290 & $41.1^{+9.4}_{-7.1}$ & $2.89^{+0.67}_{-0.63}$ &  187/177 & $48.8^{+11.5}_{-8.6}$ & $0.93^{+0.31}_{-0.26}$ &  187/177 & 1.08\\ 
[1ex]
NGC 5506 & $159.7^{+123.3}_{-36.7}$ & $0.55^{+0.13}_{-0.3}$ &  1937.51/1825 & $204.8^{+111.7}_{-53.0}$ & $0.09^{+0.04}_{-0.05}$ &  1937.34/1825 & 12.37\\ 
[1ex]
NGC 7213 & $32.1^{+32.9}_{-13.0}$ & $3.11^{+1.44}_{-1.47}$ &  914/876 & $34.1^{+40.4}_{-14.0}$ & $1.15^{+0.71}_{-0.68}$ & 914/869 & 2.64\\ 
[1ex]
% Pictor A & $77.9^{+122.8}_{-19.4}$ & $1.58^{+0.43}_{-0.85}$ &  745.16/748 & $79.2^{+49.0}_{-35.3}$ & $0.52^{+0.34}_{-0.15}$ &  745.12/748 & 1.40\\ 
% [1ex]
\hline\hline
\end{tabular}
\label{tab:params}
\end{center}
\end{table*}

\begin{table}[htb!]
\begin{center}
\caption{Best-fit parameters of the modeled reflection features in the subset of our sample of LLAGN with observed iron emission lines and/or Compton hump. For a detailed discussion, refer to section~\ref{sec:analysis}.}
\begin{tabular}{ c c c c c }
\hline  & \multicolumn{2}{c}{Gauss} & \textsc{apec} & \textsc{pexmon} \\ 
\hline \\ LLAGN name	& Energy & Width & Energy & Refl frac \\ 
 & (keV) & (keV) & (keV) &  \\ \hline
M81 & $6.61^{+0.04}_{-0.04}$ & $0.15^{+0.05}_{-0.05}$ & $1.5^{+0.06}_{-0.05} $ & ... \\ 
[1ex]
M87 & $6.58^{+0.01}_{-0.03}$ & $0.0^f$ & $1.39^{+0.03}_{-0.05} $ & ... \\ 
[1ex]
NGC 1052 & $6.3^{+0.05}_{-0.06}$ & $0.16^{+0.11}_{-0.11}$ & ...  & -0.36  \\ 
[1ex]
NGC 2992 & $6.4^f$ & $0.38^{+0.1}_{-0.08}$ & ... & $-0.67$ \\ 
[1ex]
 & $6.4^f$ & $0.34^{+0.03}_{-0.03}$ & ... & $-0.7$ \\ 
[1ex]
NGC 3227 & $6.28^{+0.02}_{-0.02}$ & $0.2^{+0.04}_{-0.04}$ & ... & $-0.14$ \\ 
[1ex]
NGC 4579 & $6.49^{+0.04}_{-0.04}$ & $0.35^{+0.05}_{-0.04}$ & ... & ... \\ 
[1ex]
NGC 5033 & $6.4^f$ & $0.0^f$ & ... & $-0.86$ \\ 
[1ex]
NGC 5273 & $6.4^f$ & $0.37^{+0.08}_{-0.07}$ & ... & $-0.72$ \\ 
[1ex]
NGC 5506 & $6.3^{+0.01}_{-0.01}$ & $0.23^{+0.01}_{-0.02}$ & ... & $-1.0$ \\ 
[1ex]
NGC 7213 & $6.53^{+0.04}_{-0.04}$ & $0.33^{+0.05}_{-0.04}$ & $7.57$ & ... \\ 
[1ex]
\hline\hline
\end{tabular}
\label{tab:gauss}
\end{center}
\end{table}

\end{document}